\documentclass{article}

\usepackage{amsmath}
\usepackage{amssymb}
\usepackage{epsfig}
\usepackage{changebar}
\usepackage{cite}

\begin{document}

\title{Positive Kauzmann Temperature: A Thermodynamic Proof and Confirmation by Exact calculations}
\author{P. D. Gujrati 
\\Department of Physics, Department of Polymer Science, \\
The University of Akron, Akron, OH 44325, USA}
\date{\today}
\maketitle
\begin{abstract}
Assuming the existence of stationary metastable states (SMS's), and using
general thermodynamic arguments, we prove that a positive Kauzmann temperature
exists for SMS's provided the ideal glass energy is higher than the crystal
energy at absolute zero. We confirm the general predictions by two exact
calculations, one of which is not mean-field.
\end{abstract}

Glass transition is a ubiquitous phenomenon$^{1,2}$ which is believed to occur
in a supercooled liquid (SCL), and has been investigated for at least over
eight decades since the earliest works of Nernst and Simon$^{3}$. Despite
this, a complete understanding of the transition and related issues is still
far from complete, although major progress has been made recently.$^{4-7}$
Theoretical and experimental investigations invariably require applying
(time-independent) thermodynamics to SCL to extract quantities like the
entropy. This presumes, as we do here, that there exists a \textit{stationary
limit} (SMS) of the metastable state (SCL) under infinitely slow cooling in
which the crystal is\textit{\ }forbidden from nucleating.$^{8}$ We further
assume that SCL can be defined all the way down to absolute zero, which may
not be the case always.$^{7}$

We focus on the partition function (PF) determined by the configurational,
i.e., the potential energy and the related entropy in this work due the
central importance of the configurational entropy, a classical statistical
mechanics (CSM) concept,$^{9}$ for the ideal glass transition.$^{1,2}$
Experimentally, the SCL configurational entropy exhibits a rapid drop near a
temperature $\simeq$ two-thirds of the melting temperature $T_{\text{M}}%
$,$^{1,2}$ and its smooth low-temperature \textit{extrapolation} gives rise to
a negative entropy (the \emph{entropy crisis} commonly known as the Kauzmann
paradox)$^{1}$ below a non-zero Kauzmann temperature $T_{\text{K}}.$ There are
several computational$^{5}$ and theoretical$^{6,7,10}$ results clearly
demonstrating the existence of the entropy crisis. A positive $T_{\text{K }}%
$is also conceptually necessary for the observed Vogel-Fulcher behavior in
fragile systems. However, the situation has become very confusing, as there
have appeared several recent arguments$^{11-13}$ against its existence. It is,
therefore, extremely important to clarify the issue, which we do by proving in
two different ways the existence of a \emph{non-zero} $T_{\text{K}} $ using
general but rigorous thermodynamic arguments, valid for classical or quantum
systems. The proof assumes SMS- \textit{existence}, so that thermodynamics can
be applied. Its stationary nature allows us to investigate SMS using the PF
formalism in (1) below. We verify our conclusions by two exact calculations,
one of which is not a mean-field calculation.

Consider a system composed of $N$ particles for which $E,V$ and $S,$ which are
homogeneous functions of order 1 in $N$, denote the energy, the volume, and
the entropy. The canonical PF for fixed $V$ and $N$ is given by%

\begin{equation}
Z\equiv TrW(E)\exp(-\beta E),\tag{1}%
\end{equation}
where $Tr$ stands for summation over the energy (eigenvalues) $E\geq E_{0}$ up
to its maximum$,$ $\beta\equiv1/T,$ $T$ being the system temperature in the
units of the Boltzmann constant $k_{\text{B}}$, and $W(E)=\exp(S)\geq1$ is
defined suiatbly to represent the \emph{number} of distinct microstates for a
given $E.$ The lowest allowed energy $E_{0}$ [$W(E_{0})\neq0]$ certainly
represents the Helmholtz free energy and the energy of the perfect crystal
(CR) at $T=0$. In equilibrium, $Z$ must be \emph{maximized} in the
thermodynamic limit $N\rightarrow\infty,$ keeping $E/N,V/N$ fixed. The
entropy$^{9}$ $S(E)\equiv\ln W(E)$ must be \textit{non-negative} so that the
state is observable in Nature.$^{14}$ This \textit{observability} condition on
$S$ \textit{cannot} be violated, even for a SMS if it is observable.
Therefore, its violation ($S_{\text{SMS}}(T)<$ $0$ for $T<T_{\text{K}}%
>0$)$^{14}$ implies that SMS \emph{cannot} exist in Nature when the violation
occurs, in which case the SMS associated with the SCL must be replaced by a
new state, commonly known as the \textit{ideal glass state} below
$T_{\text{K}},$ whose energy at $T\leq T_{\text{K}}>0$ is $E_{\text{K}}>E_{0}%
$.$^{1,2}$ The transition between the two states is called the ideal glass transition.

The order parameter $\rho$\ (see also below) is traditionally used to
distinguish various phases of the system, with $\rho=0$ representing the
disordered phase, the equilibrium liquid (EL), containing the number of
microstates $W_{\text{dis}}$ consistent with $\rho=0$. The ordered phase CR
has $\rho\neq0,$ and contains the number of microstates $W_{\text{ord}}$
consistent with $\rho\neq0.$ We partition $W(B)$ using $\rho:$ $W(B)\equiv
W_{\text{ord}}(B)+W_{\text{dis}}(B)$. (Presence of other configurations will
not affect our argument.) We introduce corresponding PF $Z_{\alpha}$
($\alpha=$ dis or ord) following (1) with $W$ replaced by $W_{\alpha}$. Above
$T_{\text{M}},$ $Z_{\text{dis}}$
$>$%
$Z_{\text{ord}},$ and below $T_{\text{M}},$ $Z_{\text{dis}}$%
$<$%
$Z_{\text{ord}}$ . The PF along with the its maximization requirement allows
us to describe CR and EL collectively. We describe SMS as
the\textit{\ continuation} of $Z_{\text{dis}}$ below $T_{\text{M}}$ by
abandoning this requirement. This necessarily yields $\rho=0$ also for SCL.
While $W_{\alpha}(E)\geq1$ for physically observed states, there is no reason
for it to hold everywhere for SMS states obtained by the above continuation;
states with $W_{\alpha}(E)<1$ are not realizable.$^{14}$ Nevertheless, as long
as $W_{\alpha}(E)\geq0$, $Z_{\alpha}$ is a sum of positive terms, so that
$\Omega_{\alpha}\equiv\ln Z_{\alpha\text{ }}$(i) satisfies proper convexity
properties$^{14}$,and (ii) is dominated by the maximum value of $S_{\alpha
}-\beta E$ in $Z_{\alpha}$. Expressing
\begin{equation}
Z_{\alpha}=W_{\alpha}(E_{0})e^{-\beta E_{0}}(1+\overline{Z}_{\alpha}),\tag{2}%
\end{equation}
where $\overline{Z}_{\alpha}\equiv Tr^{^{\prime}}[W_{\alpha}(E)/W_{\alpha
}(E_{0})]e^{-\beta(E-E_{0})},$ with $Tr^{^{\prime}}$ restricted to all
$E>E_{0},$ we find that $T\Omega_{\alpha}=TS_{\alpha}(E_{0})-E_{0}%
+T\ln(1+\overline{Z}_{\alpha}).$ For all finite $N$, $\overline{Z}_{\alpha
}\rightarrow0$ as $T\rightarrow0$, provided $W_{\alpha}(E_{0})>0$. The last
condition means that $TS_{\alpha}$ $\rightarrow0$ as $T\rightarrow0;$ hence,
$F_{\alpha}\equiv-T\Omega_{\alpha}\rightarrow E_{0}$ for CR and SMS at $T=0:$
\textit{Both have identical free energies at absolute zero}, so that
$F_{\alpha}/E_{0}\equiv1.$ Now, we can take the limit $N\rightarrow\infty$
without altering the conclusion $F_{\alpha}/E_{0}\equiv1.$ However, it is
known experimentally that (1) the heat capacity of the glassy state can be
substantially different from that of the corresponding crystal at the same
temperature,$^{15}$ and (2) the energy of this state $E_{\text{K}}$
extrapolated to absolute zero is higher than that of the CR energy
$E_{\text{0}}$.$^{1,2}$ Consequently, the stationary limit of the glassy state
is different not only from the CR, but also from the SMS near absolute zero.
Since SMS has a positve heat capacity, it achieves its energy $E_{\text{K}%
}>E_{0}$ at a positive temperature $T_{\text{K}},$ where it merges with the
ideal glass. This is our first rigorous proof that an ideal glass transition
\ must occur at a positive temperature if we identify the stationary limit of
the glassy state with the ideal glass noted above. 

\begin{figure}[p]
\begin{center}
\epsfig{file=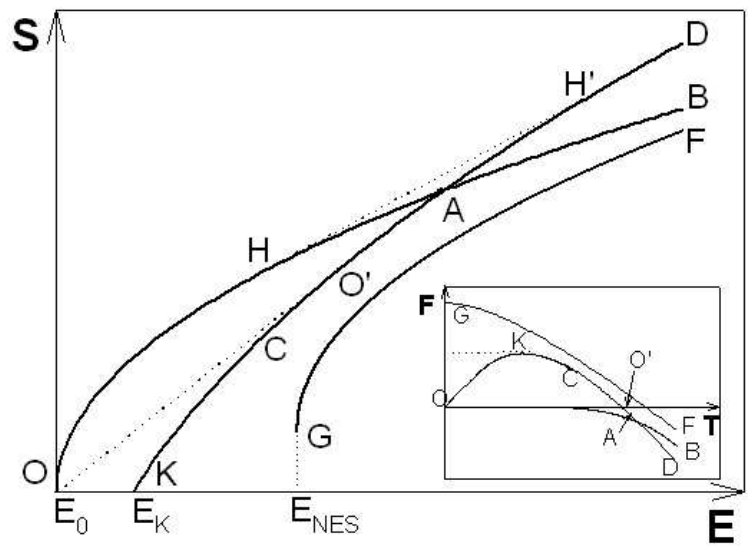,width=4.5in}
\caption{Schematic form of the generic entropy functions for various possible
states.}%
\end{center}
\end{figure}

We now proceed to give an alternative rigorous proof independent of
$F_{\alpha}/E_{0}\equiv1$ at $T=0$ derived in the canonical ensemble. We
consider the microcanonical ensemble in which the entropy functions
$S_{\alpha}$ for fixed $V$,$N$, are shown schematically as functions of $E$ by
OAB for the CR (OA) and its SMS extension (AB) for the superheated CR, and
DACK for the EL (DA), and its SCL extension (ACK) in Fig. 1. The corresponding
Helmholtz free energies $F_{\alpha}$ vs $T$ \ in the canonical ensemble are
shown schematically by BO and DACK in the inset. Since a SMS is not an
equilibrium state, its entropy at some $E$ \textit{cannot} exceed the entropy
of the corresponding equilibrium state at the same $E$. This explains the form
of the entropy in Fig. 1, which is also supported by all known
observations,$^{1,2}$ exact$^{6,7,10\text{b}}$ and numerical
\ calculations,$^{\text{5}}$ and from the arguments and the calculations
presented below. We also show the entropy function for a non-stationary
metastable state by GF in Fig. 1, assuming that crystallization is forbidden.
The slope of HH$^{^{\prime}}$ gives $1/T_{\text{M}}$ and of OO' gives the
inverse temperature (see O$^{^{\prime}}$ in the inset) where $F_{\text{SMS}%
}=E_{\text{0}}.$

Since $S,S_{\alpha}$ and $E$ are homogeneous functions of order 1 in $N,$
$(\partial S/\partial E)_{N,V}$ and its analogue $(\partial S_{\alpha
}/\partial E)_{N,V}$ are homogeneous functions of order 0 in $N$ so that
$\qquad\qquad$%
\begin{equation}
1/(\partial S/\partial E)_{N,V}=T+O(1/N),1/(\partial S_{\alpha}/\partial
E)_{N,V}=T_{\alpha}+O(1/N),\tag{3}%
\end{equation}
where $T$,$T_{\alpha}$ represent the temperatures corresponding to
$S$,$S_{\alpha}$, respectively, as $N\rightarrow\infty.$ The behavior in (3)
is critical in understanding the entropy function slope at K, the Kauzmann
point [$S_{\text{SMS}}=0]$. There are two distinct possibilities. The slope at
K is either finite, as shown explicitly in Fig. 1 and consistent with our
rigorous analysis, or infinite. The former corresponds to a positive
$T_{\text{K}},$ while the latter corresponds to an absence of a Kauzmann
paradox. Almost all explicit calculations\ show the former as the usual
behavior, two of which are presented below.

We focus on the behavior of the growth function,$^{16}$ which is adapted here
with a slight modification. We divide the almost continuum energy spectrum $E$
into bins, each of width $\Delta\epsilon>0,$ and index the bins by an integer
$j$, with $j=0$ for the lowest energy bin, which contains all energies between
$E_{0}$ and $E_{0}+\Delta\epsilon$. We may take for $\Delta\epsilon$ the
average energy difference between any two consecutive energy levels of a
single particle spectrum corresponding to its bound state; its actual choice
is not important for our discussion. Let $W(B)$ denote the number of distinct
configurations corresponding to energies lying in the bin with index $B$. The
growth function$^{16}$ $\overline{\mu}(B)$ for fixed $N,V$ is given by
$\overline{\mu}(B)\equiv W(B+1)/W(B),$ and represents on an average the number
of new configuration each older configuration generates as we add a quantum of
energy to the system. It is easy to see that $\overline{\mu}(B)$ is a
homogeneous function of order 0 in $N$. Taking the logarithm on both sides and
expanding $S(B+1)$ about $B$ for finite $N$, we immediately recognize that%

\begin{equation}
\ln\overline{\mu}(B)=(\partial S/\partial B)+O(1/N)=\Delta\epsilon
/T+O(1/N)\geq0\tag{4}%
\end{equation}
for non-negative $T,$ where we have used (3). We immediately note that not
only $\overline{\mu}(B)\geq1,$ but also that it is monotone decreasing
function of $B$. It achieves its maximum value $\overline{\mu}(B)\rightarrow
\infty$ at $T$=0, see point O in Fig. 1, and becomes \emph{finite} as soon as
$T$ is raised above zero, decreasing monotonically to 1 at infinite
temperatures. We similarly introduce the growth functions $\overline{\mu
}_{\alpha}$ using $W_{\alpha},$ which satisfy Eq. (4) except that $T$ is
replaced by $T_{\alpha}.$ To proceed further, we write $W(B)\equiv
W_{\text{ord}}(B)[1+X_{\text{dis}}(B)/X_{\text{ord}}(B)],$ where $X_{\alpha
}\equiv W_{\alpha}/W$ ($0\leq X_{\alpha}\leq1$). It is easy to see that
\begin{equation}
\ln\overline{\mu}\equiv\ln\overline{\mu}_{\text{ord}}+\ln[1+X_{\text{dis}%
}(\overline{\mu}_{\text{dis}}/\overline{\mu}_{\text{ord}}-1)]\text{.}\tag{5}%
\end{equation}
We emphasize that $\ln\overline{\mu}$ and $\ln\overline{\mu}_{\alpha}$ take
their limiting values $\Delta\epsilon/T,$ and $\Delta\epsilon/T_{\alpha},$
respectively, as $N\rightarrow\infty$; however, even for finite but large $N$,
they are not far from their limits as they are homogeneous functions of order
0. For finite $N$, $X_{\text{dis}}$ is strictly non-zero but \emph{vanishes}
as $N\rightarrow\infty$. Let us apply (5) at $B=B_{\text{K}}$, which
corresponds to the bin containing $E=E_{\text{K}}>E_{0}$. Since CR is the
equilibrium state at $E_{\text{K}},$ we have, for $N\rightarrow\infty$,
$\ln\overline{\mu}$ $\equiv\ln\overline{\mu}_{\text{ord}}=\Delta
\epsilon/T_{\text{ord}},$ the CR growth function (at $E_{\text{K}})$, which
must be finite since $E_{\text{K}}$ corresponds to a non-zero CR temperature.
The effect of $\overline{\mu}_{\text{dis}}$ is noticed only for finite $N$ as
we discuss below. For finite $N$, $\ln\overline{\mu}$ must be close to
$\Delta\epsilon/T_{\text{ord}}$. This immediately implies that $\overline{\mu
}_{\text{dis}}$ cannot diverge at $E_{\text{K}}.$ It must be \emph{finite},
and must remain finite even as $N\rightarrow\infty.$ For this to be so, it is
clear that $T_{\text{dis}}\equiv T_{\text{K}}$ cannot be zero at $E_{\text{K}%
}$. Thus, as long as $E_{\text{K}}>E_{0}$, $\overline{\mu}_{\text{dis}}$ must
remain finite as $N\rightarrow\infty$, and\ there must exist a $positive$
Kauzmann temperature $T_{\text{K}}$. This constitutes the second rigorous
proof of our central result. In CSM, $S$ represents the classical
configurational entropy.$^{9}$ In QSM, $S$ represents the total entropy.$^{9}$
Thus, $T_{\text{K}}$ corresponds to vanishing of different entropies in the
two cases. Any attempt to estimate the classical configurational entropy in
QSM, where it has no meaning, will require some sort of approximation, which
we do not consider here as we are interested in rigorous results. Therefore,
we will not pursue this point further. \qquad\qquad\qquad\qquad\qquad
\qquad\qquad\qquad\qquad\qquad\qquad\qquad\qquad\qquad\qquad\ \ \ \ \ 

Since the minimum energy state $E=E_{\text{K}}$ in SMS corresponds to a
positive temperature, the SMS state cannot exist below it. Hence, an ideal
glass transition must be invoked at this Kauzmann point ($T=T_{\text{K}})$.
The discussion also establishes that the Kauzmann point in the disordered
phase exists only when there exists another equilibrium state; otherwise,
there will be no partitioning and, therefore, there will be no Kauzmann point;
see also the explicit calculation below.

We consider two CSM models in which we obtain positive Kauzmann temperature.
The calculations are carried out exactly

1. One-dimensional Polymer Model. The example also shows how $\rho$ is used to
distinguish different phases. Consider $m$-component spins $\mathbf{S}_{i}$
located at site $i$ of a one-dimensional lattice with periodic boundary
condition ($\mathbf{S}_{N+1}$ =$\mathbf{S}_{1}$ ). Each spin can point along
or against the axes (labeled $1\leq k\leq m$) of an $m$-dimensional spin space
and is of length $\sqrt{m}$. The spins interact via a ferromagnetic
nearest-neighbor interaction of strength ($-J$), with $K\equiv J/T>0.$ The
high-temperature expansion of the PF is%

\begin{equation}
Z_{N}(K,m)=\sum K^{B}m^{L},\tag{6}%
\end{equation}
where $m\geq0$ can be taken as a real number, and describes polymers with
multiple bonds and loops, with $K$, and $m$ denoting the bond and the loop
activity, respectively.$^{17}$ The transfer matrix $\exp(K\mathbf{S\cdot
S}^{\prime})$ has eigenvalues $[x\equiv\exp(Km)]$
\begin{equation}
\lambda_{1}=x+1/x+2(m-1),\;\lambda_{2}=x-1/x,\;\lambda_{3}=x+1/x-2,\tag{7}%
\end{equation}
that are 1-fold, $m$-fold, and ($m-1$)-fold, respectively.$^{18}$ The
eigenvalue $\lambda_{1}$ is dominant at high temperatures and describes the
disordered phase. Its eigenvector $\left\langle \xi_{\text{dis}}\right|
=\sum_{i}$ $\left\langle i\right|  /\sqrt{2m},$ where $\left\langle 2k\right|
$ ( $\left\langle 2k+1\right|  $) denotes the spin state pointing along the
positive (negative) $k$-th spin-axis, has the correct symmetry to give zero
magnetization ($\rho=0$). For $m<1,$ $\lambda_{2}$ becomes dominant\ at low
temperatures [$x\geq x_{\text{c}}=1/(1-m)]$ and describes the ordered phase$,$
with a phase transition at $x_{\text{c}}$. The corresponding $m$ eigenvectors
are given by the combination $\left\langle \xi_{\text{ord}}^{(k)}\right|
=[\left\langle 2k\right|  $ $-\left\langle 2k+1\right|  ]/\sqrt{2}$, which are
orthogonal to $\left\langle \xi_{\text{dis}}\right|  .$ These eigenvectors
have the symmetry to ensure $\rho\neq0$. We note that both eigenvalues give
the \emph{same} free energy again at absolute zero. This means that if the
eigenvalue $\lambda_{1}$ is taken to represent the metastable phase above
$x_{\text{c}}$, its free energy must become equal to that of the equilibrium
phase (given by $\lambda_{2}$) at absolute zero, in conformity with the proof
given above. The bond and the loop densities $\phi_{B}\equiv$ $\partial
\ln\lambda_{1}/\partial\ln K$, and $\phi_{L}\equiv$ $\partial\ln\lambda
_{1}/\partial\ln m$ are used to calculate the entropy per site $s=\ln
\lambda_{1}-\phi_{B}\ln K-$ $\phi_{L}\ln m.$ The proper convexity requirements
($\partial\phi_{B}/$ $\partial\ln K\geq0,\partial\phi_{L}/$ $\partial\ln
m\geq0$) for (6) are always satisfied for $\lambda_{1};$ see, for example, the
behavior of $\phi_{B}$ in Fig. 2, where we have taken $m=0.7$. As the
high-temperature disordered phase represents a physical system, it cannot give
rise to a negative entropy; however, its metastable extension violates it as
shown in Fig. 2, where its entropy becomes negative below $T_{\text{K}}%
\cong0.266,$ which is lower than the transition temperature $T_{\text{c}}.$ As
$m\rightarrow0$, both $T_{\text{K}}$ and $T_{\text{c}}$ ($T_{\text{K}%
}<T_{\text{c}})$ move down to zero simultaneously, and there is \emph{no}
Kauzmann point since there is no ordered state any more, as argued above.
Thus, our exact calculation confirms our results.

\begin{figure}[p]
\begin{center}
\epsfig{file=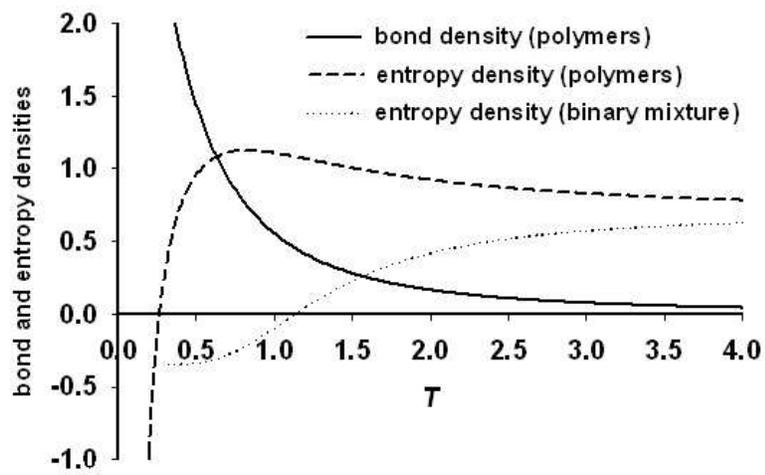,width=4.5in}
\caption{The bond and the entropy densities. Both models show an entropy
crisis at a positive temperature.}%
\end{center}
\end{figure}

It should be noted that the eigenvalues $\lambda_{1}$ and $\lambda_{2}$ are
independent of the size of the lattice. Therefore, they can be used to
describe not only the disordered and ordered phases, but also the SMS, which
is the continuation of the disordered phase, even for a finite $N$.
Thermodynamic limit is not necessary. For finite $N$, $Z^{(\alpha
)}(K,m)<Z(K,m)$ and the inequality becomes an equality only as $N\rightarrow
\infty$ for the proper choice of $\alpha$ depending on the temperature.

2. Binary Mixture Model. We now consider an anti-ferromagnetic Ising model
($J>0$) in zero magnetic field on a Husimi cactus, which can be solved
exactly.$^{19}$ The cactus can be thought as an approximation of a square
lattice, so that the exact Husimi cactus solution can be thought of as an
approximate solution of the square lattice model. The two spin states
represent the particles of two species. There is a sublattice structure at low
temperatures caused by the anti-ferromagnetic interaction:\ particles of one
species are found on one of the two sublattices. We identify this ordered
structure as a crystal. We allow two-, and three-spin ($J^{\prime}\neq0$)
interactions within the same square; the latter ensures that the melting
transition is first order. The interaction energy is%

\begin{equation}
\mathcal{E=}J\sum SS^{\prime}+J^{\prime}\sum SS^{\prime}S^{\prime\prime
}.\tag{8}%
\end{equation}
The first sum is over nearest-neighbor spin pairs and the second over
neighboring spin triplets within each square. We set $J$=1 to set the
temperature scale. The model is solved recursively, as has been described
elsewhere.$^{19}$ We introduce partial PFs $Z_{m}(\uparrow$ or $\downarrow) $
depending on the states of the spin at the $m$-th cactus level. It represents
the contribution of the part of the cactus above that level to the total PF.
We then introduce the ratio $x_{m}\equiv Z_{m}(\uparrow)/[Z_{m}(\uparrow
)+Z_{m}(\downarrow)],$which satisfies the recursion relation \begin{sloppypar}$x_{m}\equiv
f(x_{m+1},x_{m+2},v)/[f(x_{m+1},x_{m+2},v)+f(y_{m+1},y_{m+2},1/v)],$ where
$f(x,x^{\prime},v)\equiv x^{2}x^{\prime}/u^{4}v^{4}+2xx^{\prime}yv^{2}%
+x^{2}yv^{2}+u^{4}x^{\prime}y^{2}+2xyy^{\prime}+y^{2}y^{\prime}/v^{2}$ with
$u\equiv e^{\beta},v\equiv e^{\beta J^{\prime}},y\equiv1-x$ and $y^{\prime
}\equiv1-x^{\prime}$. There are two kinds of fix-point solutions of the
recursion relation, which describe the bulk behavior.$^{19}$ \end{sloppypar}In the 1-cycle
solution, $x_{m}=1/2$ is the same at all levels and exists at all
temperatures. This solution corresponds to the disordered paramagnetic phase
at high temperatures and the SMS below the melting transition. In the 2-cycle
solution, $x_{m}$ alternates between two consecutive sites. This solution
corresponds to the low temperature AF-ordered phase, which represents the CR
at half occupation. For $J^{\prime}=0.01,$ we find that $T_{\text{M}}%
\cong2.753$, where there is a discontinuity in the entropy per site of 0.0168.
The SMS below $T_{\text{M}}$ represents SCL, whose entropy density, see Fig. 2
vanishes at $T_{\text{K}}\cong1.132,$ and whose specific heat (not shown)
remains positive with a maximum at $T\cong1.26$. At absolute zero, the entropy
per site $S_{\text{SCL}}\simeq-0.3466.$ The CR and SCL free energies per site
become identical ($=2J$) at absolute zero in accordance with the claim above.
Thus, the free energy diagram we obtain in this case is similar to that in
Fig. 1.

Note that the one-dimensional exact calculation is not a mean-field type
calculation. Hence, our results are not of mean-field nature. We also note
that there is no frustration in the Ising model, which shows the ideal glass
transition. Both calculations confirm the general predictions. In particular,
we conclude that if SMS exists, it gives rise to a positive Kauzmann
temperature provided the ideal glass energy $E_{\text{K}}>E_{0},$ the ideal CR\ energy.

\begin{center}
{\large References}
\end{center}

\begin{enumerate}
\item  W. Kauzmann, Chem. Rev., \textbf{43}, 219-256 (1948).

\item \textit{The glass transition and the nature of the glassy state}, M.
Goldstein and R. Simha, eds. Ann. N. Y. Acad. Sci. 279 (1976).

\begin{sloppypar}
\item  W. Nernst, \textit{Die theoretischen und praktischen Grundlagen des
neuen W}$\overset{..}{a}$\textit{rmes}$\overset{..}{a}$\textit{tzes}$,$ Halle
(1918). F. Simon, \textit{Ergebnisse der exacten Naturwissenschaften,}%
\textbf{\ 9}, 222 (1930); 40-th Gutrie Lecture, Year Book Physical Society
(London), \textbf{1}, (1956).
\end{sloppypar}

\item  F. H. Stillinger, Science, \textbf{267}, 1935 (1995). L. Santen and W.
Krauth, Nature \textbf{405}, 550 (2000). S. Sastry, Nature, \textbf{409}, 164 (2001).

\item  F. Sciortino, W. Kob, and P. Tartaglia, Phys. Rev. Lett. \textbf{83},
3214 (1999). B. Coluzzi, G. Parisi, and P. Verrocchio, Phys. Rev. Lett.
\textbf{84}, 306 (2000).

\item  P. D. Gujrati and A. Corsi, Phys. Rev. Lett. \textbf{87}, 025701
(2001); A. Corsi and P. D. Gujrati, Phys. Rev. E \textbf{68}, 031502 (2003), cond-mat/0308555.

\item  P. D. Gujrati, S. S. Rane and A. Corsi, Phys. Rev E \textbf{67}, 052501(2003).

\item  P. G. Debenedetti, \textit{Metastable liquids:Concepts and Principles,
}Princeton University Press, Princeton, New Jersey\textit{\ (1996).}

\item  In general, the total PF can be reduced to a product\ of different PF's
corresponding to independent degrees of freedom (DOF's). Collisions between
particles, which are not included in the PF, or weak quantum effects bring
about a common temperature between these DOF's; otherwise their temperatures
need not be the same. The factorization implies that the entropy contribution
due to each DOF must not only be additive, but also non-negative individually
to avoid the entropy crisis. We only consider the irreduible PF in (1)
containig the interaction energy in the rest of the paper. In quantum
statistical mechanics (QSM) for a system at rest, the total Hamiltonian
including the kinetic energy operator determines the PF in (1); thus, $S$
represents the total entropy $S_{\text{tot}}$ including the kinetic energy
contribution. In classical statistical mechanics (CSM), the translational and
configurational PF's with corresponding entropies $S_{\text{KE}}$ and $S$,
respectively, are completely decoupled, with (1) describing the latter PF. In
this case, $S$ is the classical configurational entropy and the total entropy
is $S_{\text{tot}}=$ $S+S_{\text{KE}}.$ Although it is not the convention, $S$
in QSM will be called the quantum configurational entropy.

\item (a) J. H. Gibbs, and E. A. DiMarzio, J. Chem. Phys. \textbf{28}, 373
(1958). (b) B. Derrida, Phys. Rev. B\textbf{\ 24}, 2613 (1981).

\item  F. H. Stillinger, J. Chem. Phys. 88, 7818 (1988).

\item  D. Kivelson, S. A. Kivelson, X. Zhao, Z. Nussinov, and G. Tarjus,
Physica A 219, 27 (1995).

\item  G. P. Johari, J. Chem. Phys. 113, 751 (2000).

\item  The stability criteria such as a non-negative heat capacity that
immediately follow from the PF\ formulation are independent of the
non-negative entropy requirement. Thus, it is possible for the SMS to have a
negative entropy over some temperature range. This will only means that such
states are not observable in Nature.

\item  P. D. Gujrati and M. Goldstein, J. Phys. Chem., \textbf{84}, 859
(1980), and references therein.

\item  P. D. Gujrati, Phys. Rev. E \textbf{51}, 957 (1995).

\item  P. G. de Gennes, Phys. Lett. \textbf{38 A}, 339 (1972). P. D. Gujrati,
Phys. Rev. A \textbf{38}, 5840 (1988).

\item  P. D. Gujrati, Phys. Rev. B \textbf{32}, 3319 (1985).

\item  P. D. Gujrati, Phys. Rev. Lett. \textbf{74}, 809 (1995).
\end{enumerate}
\end{document}